\documentstyle[preprint,aps,eqsecnum]{revtex}
\begin{document}
\draft
\title{Interacting electrons on a half line coupled to impurities}
\author{Ana Lopez$^1$ , Manuel Fuentes$^{1,2}$ and Eduardo Fradkin$^2$}
\address{
$^1$ Department of Physics , Theoretical Physics , Oxford University , 
1 Keble Rd. ,Oxford , OX1 3NP , UK.\\
$^2$Department of Physics, University of Illinois at
Urbana-\-Champaign , 1110 West Green Street,Urbana, IL 61801-3080 , USA.}
\bigskip
\maketitle

\begin{abstract}
We generalize the bosonization methods for systems in the half line
that we discussed elsewhere, to study the effects of interactions on
electronic systems coupled to impurities. 
We introduce a model for a quantum wire coupled with a quantum dot by a 
purely capacitive interaction. We show that the effect on the 
quantum wire is a dynamical boundary condition which reflects the charge
fluctuations of the quantum dot. Special attention is given
to  the role of the fermion boundary conditions and their interplay
with the electron interactions on the bosonized effective theory.
\end{abstract}
\bigskip

\pacs{PACS numbers: 72.10.Fk,72.15.Qm, 73.20.Dx}

\narrowtext

%%%%%%%%%%%%%
\def\slp{{\raise.15ex\hbox{$/$}\kern-.57em\hbox{$\partial$}}}
%%%%%%%%%%%%%
%%%%%%%%%%%%%%%%%%%%%
\def\lnA{\raise.15ex\hbox{$/$}\kern-.57em\hbox{$A$}}
\def\slD{\raise.15ex\hbox{$/$}\kern-.57em\hbox{$D$}}
\def\slB{\raise.15ex\hbox{$/$}\kern-.57em\hbox{$B$}}
\def\sls{\raise.15ex\hbox{$/$}\kern-.57em\hbox{$s$}}
\def\slbarA{\raise.15ex\hbox{$/$}\kern-.57em\hbox{$\bar A$}}
\def\sla{\raise.15ex\hbox{$/$}\kern-.57em\hbox{$a$}}
\def\slb{\raise.15ex\hbox{$/$}\kern-.57em\hbox{$b$}}
\def\slbara{\raise.15ex\hbox{$/$}\kern-.57em\hbox{$\bar a$}}
\def\slimpA{\raise.15ex\hbox{$/$}\kern-.57em\hbox{$A^{\rm imp}$}}
%%%%%%%%%%%%%%%%%%%%
%%

\section{Introduction}

The Luttinger model describes the most important qualitative features
of  the low energy phenomena of the one-dimensional electron gases in
quantum wires. This physical description is a very powerful tool in 
situations
in which the wire is effectively infinite or cases in which external
gates give rise to constrictions which effectively split the wire in two
semi-infinite systems. This picture has been exploited with great
success by Kane and Fisher\cite{kane1}. Using a renormalization group
analysis, they  showed that for repulsive interactions, the scattering
potential due to an isolated impurity scales to infinity in the low
energy regime and the system is effectively cut into two different
semi-infinite systems. In other words, the problem turns out to be
equivalent to an open boundary problem.
There are other reasons why one-dimensional systems with open boundary
conditions have drawn much attention lately. From the theoretical point
of view, it is interesting to check that as in higher dimensional
problems, the boundary critical exponents are different from the bulk
ones. From an experimental point of view, the advances in the
experimental techniques have allowed the fabrication of quantum wires
in which, due to the small dimensions, boundary effects and the
coupling of the system to the external probes can be important in the
determination of the physical properties \cite{muw,ejm}.
In the last few years, as well as a combination of bosonization and  
renormalization group techniques \cite{kane1}, the method of conformal
field theory applied to the Bethe answantz soluble models has been
used to deal with systems with open boundary conditions \cite{wbp}.

In two previous papers\cite{paper1,masa}, we considered a non-interacting
system
of fermions on a half line interacting with both boundary degrees of freedom
and with impurities. The fermions coupled to the boundary degrees of freedom
through dynamical boundary conditions, and to the impurities through
local forward and backward scattering terms in the hamiltonian.
In this paper we generalize the results obtained in \cite{paper1} to
the case of interacting fermions living in the half line coupled to an impurity
at the boundary. We consider a Luttinger liquid coupled to a
boundary degree of freedom $\theta$ through the boundary condition
$\psi_R=-e^{2i\theta}\psi_L$ as well as to forward scattering
processes. We will show that this model gives a good physical
picture of a quantum wire coupled with a quantum dot by a purely capacitive
interaction. The fluctuations of the degree of freedom $\theta$ will
turn  out to be determined by the fluctuations of the charge on the 
quantum dot.
We will also show that the bosonized currents are modified by the
electron-electron interactions as well as by the presence of the
boundary degree of freedom, $\theta$. We calculate the one-particle
Green's function in two cases, when $\theta$ is a classical degree of
freedom and when it is a dynamical variable. We study the values of
this Green's function far away and close to the boundary. As expected,
we see that the exponents describing critical phenomena at the
surfaces differ from those of the bulk.

This paper is organized as follows. In section~\ref{sec:boundary} we
introduce a physically intuitive model of a quantum wire coupled to a 
quantum dot and derive the boundary condition. In
section~\ref{sec:Fermions}  we derive the effective action for the 
bosonic degree of freedom and use it to calculate the form of 
bosonized currents in the system with a boundary degree of freedom. 
In section~\ref{sec:dynamical} we introduce a model that describes the
dynamics of the boundary degrees of freedom. In
section~\ref{sec:response} we use these results to calculate the current
correlation functions both for a system with classical and quantum
mechanical boundary degrees of freedom. In section~\ref{sec:operators}
we calculate the one particle Green's function and discuss its
asymptotic limits. 
Section~\ref{sec:conclusions} is devoted to the conclusions.
In Appendix \ref{sec:AA}  we give details of the Green function of 
fermions with singular gauge fields.

\section{Boundary Degrees of Freedom in Luttinger Liquids}
\label{sec:boundary}

Consider a quantum wire coupled to a quantum dot. For all practical
purposes we will regard the dot as a gate with charge $Q$ confined in a
region of linear size $\ell$. The charge $Q$ of the dot can change due to
fluctuations caused, for example, by a nearby reservoir (such as the tip of a
scanning
tunneling microscope). Thus, $Q$ can in principle vary in time and, in the
appropriate
circumstances, it can also be regarded as a quantum mechanical degree of
freedom. In general $Q$ will fluctuate on time scales that are
usually longer that those of the fermions in the wire.
We will assume that the dot is sitting on top of the wire
and that it is close enough to be electrically coupled to it but far enough
that there is no electron tunneling between dot and wire. Under these
circumstances, the electrons on the wire will see the dot as a gate.
We will show in this section that the
qualitative effects of charge fluctuations on the dot can be represented
by a boundary condition for the electrons on the wire. The boundary
condition will be parametrized by a phase shift angle $\theta$. We
will see that  the
fluctuations of the charge cause fluctuations of the phase shift
$\theta$.

A simple model for electrons moving in a wire interacting with a gate
consists of a system of (for the time being free) fermions moving in the
potential of the gate. We will take this potential to be a step function
of height $V_0$ acting to the left of $x=0$. The Fermi energy will be assumed
to be well below
the step, $E_F \leq V_0$. Thus, the two important parameters are
the step height measured from the Fermi energy, $\Delta V=V_0-E_F$,
and the Fermi energy $E_F$ itself.
Clearly, $\Delta V$ controls the amplitude for backscattering
whereas $E_F$ controls the amplitude for forward scattering.
An elementary quantum mechanical calculation shows that the
{\it reflection} (or backscattering) amplitude $R(p_F)$ at the Fermi wavevector
$p_F$ is given in terms of the phase shift $\theta$
\begin{equation}
R(p_F)=-e^{-2 i \theta(p_F)}
\label{eq:R}
\end{equation}
where
\begin{equation}
\tan(2 \theta(p_F))=\sqrt{{\frac{E_F}{\Delta V}}}
\label{eq:thetapF}
\end{equation}
Thus, changes in the local chemical potential determine the changes in
the phase shift. We will show that this argument can be turned into a
boundary condition for the fermions in a Luttinger model.

Let $\Psi(x)$ be the Fermi field for the electrons. The low energy physics
of the interacting one-dimensional Fermi system is described by
the right and left moving components of $\Psi(x)$ in the standard
form~\cite{emery}
\begin{equation}
\Psi(x)= e^{i p_F x} R(x)+e^{-i p_F x} L(x)
\label{eq:psi}
\end{equation}
The Fermi energy is $E_F$ and the Fermi velocity is $v_F$. We will set
the zero of the energy at the Fermi energy.
Let us denote by $A_0(x)$ and $M(x)v_F^2$ the forward scattering and the
backscattering
amplitudes of the potential generated by the dot seen by the electrons 
on the
wire. In what follows
we will assume that $A_0(x)$ varies slowly in the region of the dot and
that it vanishes sharply away from it. Similarly, we will assume that the
backscattering
amplitude $M(x)$ is very large and slowly varying close to the dot and
that it also drops to zero rapidly away from it. We will assume that 
$M(x)$
and $\ell$ are large enough so that the wire has effectively split in
two semi-infinite pieces with no direct tunneling between them. Hence we
will consider the interaction of just one half of the wire with the dot.
However,if the charge $Q$ of the dot is regarded as a quantum mechanical
degree of freedom both pieces of the wire will see each other through
the fluctuations of $Q$.

In addition to the coupling to a quantum dot, a quantum wire usually has
imperfections in shape. It can also be coupled to different sorts of
impurities, both static and dynamical, which will generally cause the 
fermions to scatter.
Examples of such impurites are two-level systems. Notice that a weakly 
coupled quantum dot interacting with the electrons somewhere along the 
wire can also be regarded as a quantum impurity.
We will model the interactions between the electrons and these impurites
(quantum mechanical or not) in terms of the forward and backward
scattering amplitudes for the electrons. In particular, the forward 
scattering
amplitude terms are
\begin{equation}
{\cal H}_{\rm forward}=A_0^{\rm imp}(x) (R^\dagger(x) R(x)+
L^\dagger(x)L(x))
\label{eq:forward}
\end{equation}
while the backward scattering amplitude terms will be given by
\begin{equation}
{\cal H}_{\rm backward}= M^{\rm imp}(x) v_F^2  (R^\dagger(x)L(x) +
L^\dagger(x)R(x))
\label{eq:backward}
\end{equation}
Impurities of this type will be discussed in the next sections. 
In this section
we will consider only the effects of the edge coupled to the quantum dot.

For a system of non-interacting fermions, the effective Hamiltonian 
(density) for the
low-energy excitations of the fermions can be written in the form
(in units of $\hbar =1$)
\begin{eqnarray}
{\cal H}=&-&v_F R^\dagger(x) i \partial_x R(x)+ 
v_F L^\dagger(x) i \partial_x L(x)
\nonumber \\
&+& A_0(x) (R^\dagger(x) R(x)+L^\dagger(x)L(x))+  
M(x) v_F^2  (R^\dagger(x)L(x) + L^\dagger(x)R(x))
\label{eq:calH}
\end{eqnarray}
Clearly, this is effectively a system of free Dirac fermions in $1+1$
dimensions. Here $v_F$ is the Fermi velocity of the electrons and the
forward and backward scattering amplitudes $A_0(x)$ and $M(x)$ can be
regarded as a time component (scalar) of a vector potential and a
(varying) mass respectively. For a quantum dot we choose the functions
$A_0^{\rm dot}(x)$ and $M^{\rm dot}(x)$ to have the form
\begin{eqnarray}
A_0^{\rm dot}(x)&=&A_0 \Theta(-x) \\
M^{\rm dot}(x)&=&M \Theta(-x)
\label{eq:potentials}
\end{eqnarray}
where $\Theta(x)$ is the step function. The amplitudes $A_0$ and $M$ are
determined by the properties of the dot, such as the charge $Q$, and 
are slowly
varying
functions of the time $t$. We will take $M$ to be a constant and $A_0$
to decay very slowly, with a characteristic scale $\ell$, the linear
size of the dot.

The one-particle states of energy $E$ of this system obey the eigenvalue
equation
\begin{eqnarray}
E R(x)&=&-i v_F \partial_x R(x)+ A_0(x) R(x)+ M(x) v_F^2 L(x) \nonumber \\
E L(x)&=&+i v_F \partial_x L(x)+ A_0(x) L(x)+ M(x) v_F^2 R(x)
\label{eq:onepart}
\end{eqnarray}
For potentials $A_0(x)$ and $M(x)$ of the form of
Eq.~(\ref{eq:potentials}), and for $M$ very large, the solution of
Eq.~(\ref{eq:onepart})
is straightforward. For $x>0$, where $A_0=M=0$, the solutions have 
eigenvalue
$E=p v_F$ and
are just plane waves
\begin{eqnarray}
{\tilde R}(x)&=&  R_+ \; e^{+i p x}\nonumber \\
{\tilde L}(x)&=&  L_+ \; e^{-i p x}
\label{eq:amps1}
\end{eqnarray}
For $x<0$, the solutions with $|E|<M v_F^2$ decay exponentially fast 
as $x \to -\infty$.
Thus we write
\begin{eqnarray}
{\tilde R}(x)&=& R_- \; e^{\kappa x}\nonumber \\
{\tilde L}(x)&=& L_- \; e^{\kappa x}
\label{eq:amps2}
\end{eqnarray}
The allowed value of $\kappa$ is
\begin{equation}
\kappa=\sqrt{(M v_F)^2-({\frac{A_0}{v_F}}-p)^2}
\label{eq:ev}
\end{equation}
while the amplitudes $R_-$ and $L_-$ satisfy
\begin{equation}
{\frac{R_-}{L_-}}=
-{\frac{M v_F^2}{A_0-pv_F-i\sqrt{(M v_F^2)^2-(A_0-pv_F)^2}}}
\label{eq:amps3}
\end{equation}
Since continuity of the wave function at $x=0$ demands that
${\frac{R_+}{L_+}}={\frac{R_-}{L_-}}$, we find that the phase shift
$\theta(p)$, defined by
\begin{equation}
{\frac{R_+}{L_+}}=-e^{2 i\theta(p)}
\label{eq:pm}
\end{equation}
is given by
\begin{equation}
\tan (2\theta(p))={\rm sgn} (A_0-p v_F) \sqrt{({\frac{M v_F^2}{A_0-p
v_F}})^2-1}
\label{eq:ps}
\end{equation}
Here we are only interested in the limit of small momentum $|p| \ll  {\rm
inf}(M v_F,A_0/v_F)$,
and potentials $|A_0| <  M v_F^2$. In this regime the phase shift becomes
\begin{equation}
\tan (2\theta(0))={\rm sgn}(A_0) \sqrt{({\frac{M v_F^2}{A_0}})^2-1}
\label{eq:ps0}
\end{equation}
~From the point of view of the states to the right of $x=0$, if we
restrict ourselves to
states with small momentum, the presence of the dot becomes 
{\it equivalent} to the boundary condition
\begin{equation}
R(0,t)=- \;e^{2 i \theta(0)}\; L(0,t)
\label{eq:bc1}
\end{equation}
Thus, all knowledge of $A_0$ and $M$, which are the qualitatively
important parameters of the quantum dot as far as the fermions on the
wire are concerned, is expressed in the phase shift $\theta(0)$. 
Clearly, the
fermions to the right of $x=0$ know about the properties of
the dot since their wavefunctions leak-out to negative values of $x$ 
and feel
the voltage $A_0$.

Consistent with this conventions, we will choose the domain of the
phase shift to be in the interval 
$ -{\frac{\pi}{2}} < 2\theta  \leq {\frac{\pi}{2}}$. As far as this
one-particle analysis goes all the  branches of the function arctangent
lead to physically equivalent results. However, in the full second
quantized theory, the phase shift will determine a dynamical boundary 
condition for the fermions. In that situation, moving the boundary 
angle from one branch to the next is no longer physically inocuous
since it changes the total number of fermions (namely, the total
charge). This physical picture is clearly not related to the model we 
are discussing here since in the abscence of direct tunneling between 
the wire and the dot, the total charge of the wire cannot change.

However, this does not mean that there cannot be charge fluctuations
near the end of the wire. Quite to the contrary,
in physically interesting situations the quantum dot can exhibit charge
fluctuations. We can now represent phenomenologically these charge 
fluctuations
$Q(t)$ in the form of a slow time dependence of the voltage 
$A_0(t) \propto Q(t)$.
Hence, by this mechanism, the boundary condition acquires dynamics. 
We define a
{\it boundary} degree of freedom $\theta(t)$ as the (adiabatic 
extension of)
phase shift $\theta(0)$. The boundary condition for the fully second
quantized right and left moving fields is now
\begin{equation}
R(0,t)=- \;e^{2 i \theta(t)}\; L(0,t)
\label{eq:bc2}
\end{equation}
As the charge $Q(t)$ on the dot changes, so does the potential $A_0(t)$.
Consequently, the electrons on the wire are pushed further to the right 
(left)
of $x=0$ as $A_0(t)$ grows larger (smaller). Hence, a fluctuating boundary
condition causes the charge density of the fermions at the edge of the 
wire to fluctuate. Physically this means that the position of the 
physical edge of the
wire changes with time and it becomes a dynamical variable with quantum
fluctuations of its own. Hence, the fluctuations of $\theta$ ``push"
the  Luttinger liquid as a whole and, as a result, they
are insensitive to the electron-electron interactions in the liquid. In
the next sections we will show that this is precisely what happens. In
contrast, the fluctuations of a forward scattering quantum impurity in
the bulk (even those close to the edge) are affected by the 
electron-electron interactions in the liquid.

\section{Luttinger model with a localized forward scattering impurity}
\label{sec:Fermions}

In this section we give a description of a Luttinger model of
interacting electrons on a half-line, coupled to
impurities located very close to the edge of the line. We
consider two types of impurities: (a) a boundary degree of freedom 
(of the type discussed in section \ref{sec:boundary}), which couples with 
the fermions through the boundary
conditions as in Eq.~(\ref{eq:bc2}), and (b) forward and backward 
scattering impurities.
In this section the impurity will be regarded as a
classical degree of freedom. In section \ref{sec:dynamical} the 
impurity will be quantized. The
approach that we use here follows closely the methods that we developed
in references \cite{paper1} and \cite{masa} for non-interacting electrons.

We begin with the Lagrangian for the massless Luttinger model
on the half-line ({\it i.\ e.\ } $1+{1\over 2}$ dimensions).
Throughout we will work in two-dimensional euclidean space 
(imaginary time).
This means that we work on
a space-time $\Omega$ which is an infinitely long cylinder, with its 
axis along the space direction
$x_1$ and periodic (antiperiodic for fermi fields) boundary conditions 
along the imaginary time
direction $x_2$ of perimeter $T$, with $T \to \infty$. 
The boundary will be denoted by
$\partial \Omega$ and it is the set of points with $x_1=0$.

The imaginary time (euclidean) Lagrangian for the pure system (coupled to
sources) is
\begin{equation}
{\cal L}={\bar \psi} \;i  \slp \; \psi- {\frac{g^2}{2}}
\left({\bar \psi}\; \gamma_\mu \psi \right)^2+{\cal A}_\mu J_\mu
\label{eq:Lth}
\end{equation}
where ${\cal A}_\mu$ is a source (a background gauge field) which
couples to the fermion current $J_\mu={\bar \psi}\; \gamma_\mu \psi $.
We will follow the procedure discussed in section~\ref{sec:boundary} and 
impose
dynamical
boundary conditions  for the fermionic field of the form of 
Eq.~(\ref{eq:bc2}).
% In Appendix \ref{sec:BB} we show that 
In Euclidean space the boundary condition becomes 
(see reference \cite{paper1})
\begin{equation}
R(0,x_2)=-e^{2\theta(x_2)}L(0,x_2)
\label{eq:EBC}
\end{equation}
The impurity forward scattering term is
\begin{equation}
{\cal L}_{\rm imp}={\bar \psi}\; \gamma_\mu \psi \; A^{\rm imp}_\mu(x)
\label{eq:Limp}
\end{equation}
where $A^{\rm imp}_\mu$ has the form
\begin{equation}
A^{\rm imp}_\mu(x)= \delta(x_1) \; {\bar A}_\mu(x_2)
\label{eq:vectoramplitude}
\end{equation}
for systems with a (possibly time-dependent) degree of freedom localized 
at $x_1=0$.

Next we proceed to follow the usual path integral bosonization
method\cite{laplata}. First we decouple the four-fermi interaction by
means of a vector (Hubbard-Stratonovich) auxiliary field $a_\mu$. In
this form, the full euclidean action for the Luttinger model
coupled to forward-scattering impurities reads,
\begin{equation}
{\cal L}={\bar \psi} \;i  \slD \; \psi+ {\frac{a_\mu^2 }{2g^2}}
\label{eq:Lfull}
\end{equation}
where the covariant derivative $D_\mu$ is
\begin{equation}
D_\mu \equiv \partial_\mu-i\left( a_\mu+{\cal A}_\mu+A^{\rm
imp}_\mu(x)\right)
\label{eq:newcovariant}
\end{equation}
which couples the fermions to the auxiliary field $a_\mu$, to the
external source ${\cal A}_\mu$ and to the vector impurity amplitude
$A^{\rm imp}_\mu$.
The dynamics of the fermions is described by the functional integral
\begin{equation}
{\cal Z}= \int {\cal D} {\bar \psi} {\cal D} \psi {\cal D} a_\mu
\exp \left(-\int d^2x {\cal L}_F \right)
\label{eq:functint}
\end{equation}
We are interested in the properties of this system when the
fermions are restricted to move on a half-line with the impurity
located close to the edge. Since one plausible effect of electron-electron
interactions is to induce (or to screen) boundary degrees of freedom, we
must define carefully the system near the edge.
We write the external sources as a sum of bulk and boundary pieces of
the form ${\cal A}_\mu = B_\mu +{\bar b}_\mu(x_2) \delta(x_1)$. Clearly,
$b_\mu$ is the value of ${\cal A}_\mu$ at the boundary, and $B_\mu$ is its
value in the bulk. We will require $B_\mu$ to vanish at the boundary, 
{\it i.~e.\/} $B_\mu(x_1=0)=0$.  For convenience,
we define $s_\mu(x)= ( {\bar A}^{\rm imp}_\mu + {\bar b}_\mu)(x_2)
\delta(x_1)\equiv {\bar s}_\mu(x_2)\delta(x_1)$.  For technical reasons, 
it will also be convenient to define the impurity to be at finite 
(but small) distance from the edge and to let $a_\mu$  vanish at the 
edge $x_1=0$ but not at the location of the impurity.

The functional integral now reads
\begin{equation}
{\cal Z}[B_\mu,b_\mu]= \int {\cal D} {\bar \psi} {\cal D} \psi {\cal
D} a_\mu
\exp \left(-S_F ({\bar \psi},\psi,a_\mu,) \right)
\label{eq:functint0}
\end{equation}
where, after shifting $ a_\mu + B_\mu \rightarrow a_\mu$, the
action $S_F$ takes the form
\begin{equation}
S_F ({\bar \psi},\psi,a_\mu)=\int_{\Omega} d^2x \;
{\bar \psi}( i \slp + \sla +\sls)\psi + \int_{\Omega} d^2x \;
{1 \over 2g^2} (a_\mu-B_\mu)^2
\label{eq:fermiaction1}
\end{equation}
The next step in the bosonization of the system is to decouple the
fermions  from the regular gauge field $a_{\mu}$. This is accomplished 
by means of a combination of (suitably chosen) smooth, single-valued 
gauge and chiral transformations of the form
\begin{eqnarray}
\psi(x)&=&e^{i\eta(x)+\gamma_5 \phi(x)}\chi(x)\nonumber\\
{\bar \psi}(x)&=&{\bar \chi}(x) \; e^{-i\eta(x)+\gamma_5 \phi(x)}
\label{eq:chiral}
\end{eqnarray}
The fermions decouple if the vector potential $a_\mu(x)$ are smooth and
can be written in the form
\begin{equation}
a_\mu(x)=\partial_\mu \eta(x)-\epsilon_{\mu \nu} \partial_\nu \phi(x)
\label{eq:vecpot}
\end{equation}
This condition can be met everywhere except at the boundary where the
configurations have a $\delta$-function support and become singular 
and will be treated separately.
The boundary condition $a_\mu(0,x_2)=0$  is satisfied if we impose 
Dirichlet
boundary conditions on $\phi$, {\it i.~e.\/}, $\phi(0,x_2)=0$,
and  von Neumann boundary conditions on $\eta$, {\it i.~e.\/},
$\partial_1\eta(0,x_2)=0$ .

Under the change of variables of Eq.~(\ref{eq:chiral}) the integration 
measure
of the fermion path integral acquires a jacobian $J_F$ of the
form \cite{paper1,laplata}
\begin{equation}
J_F = e^{ -{1\over{2\pi}}\int_{\Omega} (\partial_{\mu}\phi)^2 +
{1\over{\pi}}\int_{\Omega} {\epsilon}_{\mu\nu} s_\mu
\partial_{\nu}\phi }
\label{eq:omegaepsilon}
\end{equation}
Hence, the fermion part of the path integral becomes
\begin{equation}
\int {\cal D} {\bar \psi} {\cal D} \psi \; 
\exp\left(-\int_{\Omega} d^2x \;
{\bar \psi}( i \slp + \sla +\sls)\psi\right)=J_F \int {\cal D} {\bar \chi}
{\cal D}
\chi \; \exp\left(-\int_{\Omega} d^2x \; 
{\bar \chi}( i \slp +\sls)\chi\right)
\label{eq:jac}
\end{equation}
~From Eq.~(\ref{eq:vecpot}), it follows that $\epsilon_{\mu\nu}
\partial_{\mu} a_{\nu}= \partial^2 \phi$. Therefore the chiral angle 
$\phi$ is given by
\begin{equation}
\phi (x) = \int_{\Omega} d^2 y \;  G_D(x,y) \epsilon_{\mu\nu}
\partial_{\mu} a_{\nu}(y)
\end{equation}
where $G_D(x,y)$ is the Dirichlet Green's function in $\Omega$ which
satisfies
\begin{equation}
 \left\{
\begin{array}{ll}
\partial ^2_x G_D(x,y)= \delta(x-y)\\
G_D(x_1,x_2;y)|_{x_1=0}=0
\end{array}
\right.
\label {eq:GD}
\end{equation}
The solution of Eq.~(\ref{eq:GD}) is
\begin{equation}
G_D(x,y) = {1\over {4\pi}} \ln { {(x_1 - y_1)^2 + (x_2 - y_2)^2 + a^2}
\over{(x_1 + y_1)^2 + (x_2 - y_2)^2 + a^2} }
\label{eq:GDE}
\end{equation}
where $a$ is a short distance cut-off.

Using these properties, the jacobian $J_F$ becomes
\begin{equation}
J_F = \exp \left( { {1\over{2\pi}}\int_{\Omega} d^2 x d^2 y \; a_{\mu}(x)
\; \Gamma_{\mu\nu}(x,y) \; a_{\nu}(y) +{1\over{\pi}}\int_{\Omega} d^2 x
d^2 y \; s_{\mu}(x)\;  \Gamma_{\mu\nu}(x,y) \; a_{\nu}(y) } \right)
\label{omegaepsilon2}
\end{equation}
where the kernel $\Gamma_{\mu \nu}(x,y)$ is
\begin{equation}
{\Gamma}_{\mu \nu}(x,y)=
\left(\partial^{(x)}_{\alpha}\partial^{(y)}_{\alpha}\delta_{\mu \nu}-
\partial^{(y)}_{\mu}\partial^{(x)}_{\nu}\right)G_D(x,y)
\label {eq:gammas}
\end{equation}
%which can be calculated using Eq.~(\ref{eq:GDE}).

After substituting these results back into
Eq.~(\ref{eq:functint0}), the partition function takes the form
\begin{eqnarray}
{\cal Z}[B_\mu,b_\mu]=&\int& {\cal D} {\bar \chi} {\cal D} \chi \;
\exp  \left( - \int_\Omega d^2x [{\bar \chi}
(i\slp+\not\!s)\chi]\right)
\int {\cal D} a_\mu\;
\exp \left(-\int_{\Omega} d^2x \;{\frac{1}{2g^2}}
(a_\mu-B_\mu)^2   \right)
 \nonumber\\
&\times& \exp \left( { {1\over{2\pi}}\int_{\Omega} d^2 x d^2 y\; a_{\mu}(x)
\Gamma_{\mu\nu}(x,y) a_{\nu}(y) +{1\over{\pi}}\int_{\Omega} d^2 x
d^2 y \; s_{\mu}(x) \Gamma_{\mu\nu}(x,y) a_{\nu}(y) } \right)
\label{eq:functint2}
\end{eqnarray}
The transformation Eq.~(\ref{eq:chiral}) does not decouple the 
singular gauge
fields ( with support at the boundary) from the
fermions. In reference\cite{paper1} we showed that the result of this path
integral is
also a fermion determinant but for operators with different boundary 
conditions
(parametrized by a phase shift) which can be computed using Forman's
method\cite{forman}.
The result is\cite{paper1}
\begin{eqnarray}
 \int {\cal D} {\bar \chi} {\cal D} \chi \;
&&\exp \left( - \int_\Omega d^2x [{\bar \chi} (i\slp+\not\!s)\chi]\right)=
\nonumber\\
=&&{\rm Det} (i\slp)_{R=-L}\;\;
\exp\left({1\over {2{\pi}}}\int dx_2 dy_2
(\theta+{\bar s}_2)(y_2)K(x_2,y_2)(\theta+{\bar s}_2)(x_2)\right)
\label{eq:fordet}
\end{eqnarray}
The kernel $K(x_2,y_2)$ is equal to
\begin{equation}
K(x_2,y_2)= \left( -{1\over{\pi}} {\frac{1}{(x_2-y_2)^2+a^2}}
+ {1\over{a}}\delta(x_2-y_2) \right).
\label{eq:kernel}
\end{equation}
where $a$ is the short distance cut-off introduced above.
It is easy to show that the exponential in Eq.~(\ref{eq:fordet})
can be written in the standard Caldeira-Leggett form \cite{caldeira}
\begin{eqnarray}
{1\over {2{\pi}}}&&\int dx_2 \int dy_2
(\theta+{\bar s}_2)(y_2)K(x_2,y_2)(\theta+{\bar s}_2)(x_2)\nonumber\\
&&=
{1\over {4{\pi}^2}}\int dx_2 \int dy_2 \;\; {\frac{\left((\theta+{\bar
s}_2)(x_2)-
(\theta+{\bar s}_2)(y_2)\right)^2}{(x_2-y_2)^2}}
\label {eq:uno}
\end{eqnarray}
The integral over the Hubbard-Stratonovich field $a_{\mu}$ in
Eq.~(\ref{eq:functint2}) can be performed and the result is
\begin{eqnarray}
\int {\cal D} a_\mu &&
\exp \left({1\over 2\pi} \int_{\Omega} d^2x \int_{\Omega} d^2y \;  
a_{\mu}(x)
[\Gamma_{\mu\nu}(x,y) - {\frac{\pi}{g^2}} \delta_{\mu\nu}
\delta^2(x-y) ] a_{\nu}(y) \right.
\nonumber\\
&+&
 \left.  {1\over{\pi}}\int_{\Omega} d^2 x \int_{\Omega} d^2 y s_{\mu}(x)
\Gamma_{\mu\nu}(x,y) a_{\nu}(y) +{1\over{g^2}}\int_{\Omega} d^2 x
 a_{\mu}(x) B_{\mu}(x) \right)
\nonumber\\
&=&
\exp \left({1\over 2}\int_{\Omega} d^2x \int_{\Omega} d^2y \;  B_{\mu}(x)
[{\frac{1}{(\pi+ g^2)}} \Gamma_{\mu\nu}(x,y) + 
{\frac{1}{ g^2}}\delta_{\mu\nu}
\delta^2(x-y)] B_{\nu}(y) \right.
\nonumber\\
&+&
\left. \int_{\Omega} d^2x \int_{\Omega} d^2 y  { {1\over{(\pi+g^2)}}
[s_{\mu}(x)
\Gamma_{\mu\nu}(x,y) B_{\nu}(y) -{g^2\over{2\pi}}
s_{\mu}(x) \Gamma_{\mu\nu}(x,y) s_{\nu}(y) ]} \right)
\label{eq:inta}
\end{eqnarray}
Here we have used that the inverse of the kernel in the quadratic term
in $a_\mu$ is
\begin{equation}
{\left( {\Gamma_{\mu\nu}(x,y) \over \pi} - {\delta_{\mu\nu}
\delta^2(x-y)\over g^2} \right)} ^{-1} = - g^2 \left( {g^2 \over
{(\pi + g^2)}}  \Gamma_{\mu\nu}(x,y) + {\delta_{\mu\nu}
\delta^2(x-y)}
\right)
\end{equation}
and that $\int d^2y \; \Gamma_{\mu\nu}(x,y) \Gamma_{\nu\alpha}(y,z)=
-\Gamma_{\mu\alpha}(x,z)$.
Substituting the results for the integrals over the fermionic fields
and the integral over $a_{\mu}$ back into Eq.~ (\ref{eq:functint2}), and
using that
\begin{equation}
\int_\Omega d^2x \int_\Omega d^2y
s_\mu(x) \Gamma^{\mu\nu}(x,y) s_\nu(y)=
\int dx_2 \; \int dy_2 \; {\bar s}_2(y_2)K(x_2,y_2){\bar s}_2(x_2)
\label{eq:relation2}
\end{equation}
we obtain the following expression for the partition function
\begin{equation}
{\cal Z}[B_\mu,b_\mu]= e^{-S_{\rm ind}({\bar A}^{\rm imp}_\mu,
{\bar b}_{\mu},\theta)}
\label{eq:Zeff}
\end{equation}
The induced action $S_{\rm ind}$ given by
\begin{eqnarray}
S_{\rm ind}({\bar A}^{\rm imp}_\mu,{\bar b}_{\mu},\theta)&=&
-{\frac{1}{2(g^2+\pi)}}
\int_\Omega\!\! d^2x\!\! \int_\Omega\!\! d^2y(B_\mu (x)+{\bar A}^{\rm
imp}_\mu+{\bar
b}_\mu){\Gamma}_{\mu \nu}(x,y) (B_\nu+{\bar A}^{\rm imp}_\nu+{\bar
b}_\nu)(y)\nonumber\\
&&-\frac{1}{2\pi}\int dx_2 \int dy_2 \;\; \theta (x_2) K(x_2,y_2) 
\theta (y_2)
\nonumber\\
&&-\frac{1}{\pi}\int dx_2 \int dy_2 \;\; \theta (x_2) K(x_2,y_2)
({\bar A}^{\rm imp}_2+{\bar b}_2)(y_2)
\label{eq:ifclass}
\end{eqnarray}
The induced  action $S_{\rm ind}({\bar A}^{\rm imp}_\mu,
{\bar b}_{\mu},\theta)$ for the boundary degrees of freedom is 
explicitly gauge invariant,
and it has a smooth dependence in the coupling constant $g$. 
It also reproduces exactly  our
results for free fermions of reference \cite{paper1}.

The induced action of Eq.~(\ref{eq:ifclass}) has a clear physical meaning.
Recall that the degrees of freedom $\theta$ and ${\bar A}_2$
represent, respectively, the dynamical boundary condition and a quantum
mechanical forward
scattering impurity close to the boundary.
In contrast to the case of free fermions \cite{paper1}, once
interactions are present,
$\theta$ and  ${\bar A}_2$ are no longer physically equivalent. 
In particular,
Eq.~(\ref{eq:ifclass}) shows that while the interactions modify the 
amplitude
of the terms that contain ${\bar A}_2$ they do not enter in the terms
that involve $\theta$. Physically, the renormalization of the amplitude of the
terms involving ${\bar A}_2$ is just the screening of
the forward scattering impurity by the fluctuations of the fermions.
Since the Luttinger model has strictly short range interactions,
screening is not complete but, instead,  it leads to a 
finite reduction of the
forward scattering amplitude. On the other hand,  
the boundary degree of fredom
$\theta$, which represents the coupling to the quantum dot, 
is ``outside" the system it is not affected by screening effects.

To obtain the bosonization formula for the full partition function, we
define a bosonic field $\omega$ which obeys Dirichlet boundary
conditions $\omega(0,x_2)=0$.
Following
the same steps as in \cite{paper1} we find that the partition function
of the Luttinger model on a half-line is equal to the following
bosonic partition function
\begin{eqnarray}
{\cal Z}[B_\mu,b_\mu]={\cal K}\;\int{\cal D} \omega &&\;\exp\left(
-{1\over {2\pi}}\int (\partial_\mu \omega)^2 + {i\over
{\pi}}\sqrt{\frac{\pi}{\pi+ g^2}}
\epsilon_{\mu\nu}\partial_\nu \omega (B_\mu (x)+{\bar A}^{\rm imp}_\mu+
{\bar
b}_\mu)\right)
\nonumber\\
&&+\left.\frac{1}{2\pi}\int dx_2 \int dy_2 \;\; 
\theta (x_2) K(x_2,y_2) \theta
(y_2)
\right.\nonumber\\
&&+\left.\frac{1}{\pi}\int dx_2 \int dy_2 \;\; \theta (x_2) K(x_2,y_2)
({\bar A}^{\rm imp}_2+{\bar b}_2)(y_2)\right)
\label{eq:pfclasica}
\end{eqnarray}
This is the bosonized form of the theory. It gives an explicit
description of the system in terms of the fluctuations of the 
Dirichlet bose
field
$\omega$. It also includes the effects of both the boundary and forward
scattering degrees
of freedom $\theta$ and ${\bar A}_2$.

~From  Eq.~(\ref{eq:pfclasica}) we can read-off the boson operators which
represent the fermion currents. In the bulk,
{\it i.~e.\/} for $x_1>0$, the current operator becomes
\begin{equation}
J_{\mu}(x)\equiv{i\over {\pi}}\sqrt{\frac{\pi}{\pi+ g^2}}
\epsilon_{\mu\nu}\partial_\nu \omega
\label{eq:bose-identity}
\end{equation}
This is the standard bosonization formula for the current.  At the
boundary we find
\begin{equation}
J_{\mu}(0,x_2)\equiv \left. {i\over {\pi}}\sqrt{\frac{\pi}{\pi+ g^2}}
\epsilon_{\mu\nu}\partial_\nu \omega \right|_{x_1=0}
+{\delta_{\mu,2}\over {\pi}} \int dy_2 K(x_2,y_2) \theta (y_2)
\label{eq:jbound}
\end{equation}
The first term is the bulk contribution of the current at the
boundary while the second term is the contribution of the boundary
degreee of freedom $\theta$ and it only affects the charge density at the
boundary. This is precisely what we anticipated in section
\ref{sec:boundary}
where we showed that
the fluctuations in the phase shift $\theta$ cause the fermions to
fluctuate in and out of the wire at the edge. This is the physical meaning of
the last term of Eq.~(\ref{eq:jbound}).

The form of the current operator in Eq.~(\ref{eq:jbound}), combined 
with the
Dirichlet boundary condition for the field $\omega$, ensures
that the space component of the
fermion current vanishes exactly at the boundary. This
identification of the charge density operator at the boundary implies that, at
the
operator level, $\theta$ contributes to the boundary charge while the 
forward scattering
impurity ${\bar A}_2$ does not. In fact, the boundary charge operator is
precisely the same one
that was found in reference \cite{paper1} for the case of free fermions.
Nevertheless, the {\it expectation value} of this operator does depend on
${\bar A}_2$ due to the fluctuations of $\omega$ and this 
dependence involves
the value of the coupling constant $g$. We will discuss this issue
in  section \ref{sec:response}, where we use the partition function
computed above to find the response of the system to an external
electromagnetic field.

\section{Dynamical Boundary Conditions and Charge Fluctuations}
\label{sec:dynamical}

So far we have treated the impurities $\theta$ and ${\bar A}_2$ as classical
degrees of
freedom. We now proceed to quantize the impurity degrees of freedom. In Section
II we showed that
the fluctuations of the charge $Q(t)$ of the quantum dot can be modelled by a
dynamical boundary
condition with phase shift $\theta(t)$. We now need to define the
dynamics of the charge fluctuations in the quantum dot.

Let $|Q \rangle$ be the ground state of the dot with charge $Q$. We will be
interested in the effects of the fluctuations of the total charge $Q$. For each
value of $Q$, the dot has a Hilbert space of states with the same charge. These
states which are closely spaced in energy, only change the charge distribution
inside the dot but, at least qualitatively, do not affect strongly the phase
shift. In what follows we will only consider the ground states for sectors with
different values of $Q$. These states are in general non-degenerate
and can be labelled just by the {\it integer} $Q$.
In the abscence of tunneling the energies of the eigenstates $|Q \rangle$ of
the dot are $U(Q)$. This energy $U(Q)$ is determined by Coulomb interactions
among the electrons in the dot and by the potentials that stabilize the dot.
Although in general this spectrum is
rather complex, it must exhibit some  general features. For $Q$ large
enough the energy $U(Q)$ should have a minimum at some value of the charge
$Q_0$ determined by the dot gate voltage. In this regime $Q_0$ is large and the
spectrum of charges effectively ranges over all the integers.

Tunneling processes between the quantum dot and a reservoir (for instance a
scanning
tunneling microscope) will cause the charge to fluctuate. Hence, the tunneling
term
of the Hamiltonian acts as a kinetic energy and its scale is parametrized by
$T$,
the tunneling amplitude. Since only one particle can tunnel at a time,
the tunneling Hamiltonian must involve an operator which maps the state
$|Q\rangle$ to the states $|Q\pm 1\rangle$. Thus, we need an operator which
translates
$Q$ by one unit at a time.
Let ${\hat \phi}$ be an operator which satisfies the commutation relation
$[{\hat Q},{\hat \phi}]=i$.
Clearly ${\hat \phi}$ is the momentum canonically conjugate to ${\hat Q}$.
 Since the
spectrum of $Q$ are the integers, the canonical momentum  ${\hat \phi}$ is an
angle. Hence, we can write the dot Hamiltonian in the form
\begin{equation}
H_{\rm dot}=-T \cos {\hat \phi}+U({\hat Q})
\label{eq:Hdot}
\end{equation}
For simplicity we will approximate $U(Q) \approx {\frac{1}{2C}} (Q-Q_0)^2$,
where $C$ is the capacitance.  We will be interested in situations in which the
capacitance is large or, alternatively, the tunneling amplitude $T$ is large.
In this regime the discreteness of $Q$ can be ignored and the spectrum of $Q$
becomes the real line. Let us define the rescaled charge operator
${\hat q}=({\hat Q}-Q_0)/\sqrt{C}$ and the associated canonical momentum ${\hat
P} = \sqrt{C} {\hat \phi}$, which satisfy the commutation relation $[{\hat q},
{\hat p}]=i$. Up to a shift in the energy, the dot hamiltonian reduces
to a simple harmonic oscillator
\begin{equation}
H_{\rm eff}={\frac{T}{2C}} {\hat p}^2+ {\frac{{\hat q}^2}{2}}
\label{eq:Heff}
\end{equation}
which describes the small charge fluctuations in the quantum dot.

However, the quantum dot couples to the fermions only through the phase shift.
Hence, what we really need is not the Hamiltonian for the charge $Q$ but
instead
the Hamiltonian for  the phase shift $\theta$. In Section II we showed that $Q$
determines $A_0$, the voltage felt by the fermions. In turn, the phase
shift $\theta$ also is determined from $A_0$ by Eq.~(\ref{eq:ps}). The
actual details of this relation are not very important (in fact, they depend
very strongly on the details of the model). What is important is that $\theta$
is only defined {\it modulo} $\pi$ (see Eq.~(\ref{eq:bc2})). Thus, $U(Q)$
becomes a {\it periodic} function of $\theta$. Also, since $\theta$ is an
angle, the momentum conjugated to ${\hat \theta}$ must be an angular
momentum operator ${\hat L}$ and it should obey the commutation
relation $[{\hat \theta},{\hat L}]=i$. The kinetic energy term
should now be given in terms of ${\hat L}$. Hence, the effective 
Hamiltonian for $\theta$ must have the form
\begin{equation}
H_{\theta}={\frac{{\hat L}^2}{2K}}+U({\hat \theta})
\label{eq:Htheta}
\end{equation}
where $K \propto 1/T$. $U({\hat \theta})$ is periodic with period $\pi$ and
it has a minimum at a value $\theta_0$ determined by the charge. The
actual parameters of this hamiltonian depend on details of the model.
It is important to note that if the capacitance  $C$ is large, the potential
$U(Q)$ will be very shallow. As a result, $U(\theta)$ not only will be
very shallow but also, it will be always different from zero.
Hence, capacitance effects will always tend to suppress the
fluctuations of the phase shift. In what follows we will consider a
model in  which $U({\hat \theta})=G \cos (2\theta)$
with a coupling constant $G \propto 1/C$. This model is explicitly
periodic in $\theta$. In reference \cite{paper1} we considered a model
of non-interacting fermions on a half-line coupled to an impurity with
this type of dynamics. The coupling to a the quantum dot described here is an
explicit construction
of this model.

There are a number of interesting cases in which the dot confinement potential
is quite steep. In this situation the level spacing is large and a continuum
approximation for $Q$ is
qualitatively incorrect. In such cases it is better to think of the system as a
two level system in which the charge can take just two values, 
$Q_\pm$ with
tunneling producing transitions between these two states. Hence, there are only two allowed values
of the angle
as well. In this case the boundary degree of freedom behaves like 
an Ising degree of
freedom. If we denote the two states by $|\pm \rangle$ the 
Hamiltonian now takes the simple form
\begin{equation}
H_{\pm}=T \sigma_1+{\frac{1}{C}}\sigma_3
\label{eq:Hising}
\end{equation}
The coupling between the fermions and the impurity still takes the same
form as before but the phase shift can take just two values. This
problem is known\cite{leggett} to be related to the Kondo problem in
metals. We will not consider this case in this paper.

In the rest of this paper we will use the approach of reference \cite{paper1}
and
use path integrals to describe the quantum mechanics of both the fermions and
of the quantum dot,
the latter being represented by the phase shift $\theta$ promoted to a full
dynamical degree of freedom at the boundary.
Using standard methods one derives the path integral for the boundary degree of
freedom $\theta$. The only subtlety here is the fact that the configuration
space of $\theta$ is the interval $(-{\frac{\pi}{2}},{\frac{\pi}{2}}]$
and {\it not} the unit circle of length $2\pi$. In other words, the histories
of the boundary degree of freedom do not include configurations with non
zero winding number, namely functions $\theta(x_2)$ which are periodic
with period $\pi$. Such configurations are important to describe
tunneling processes between de wire and the dot. In what follows we will
always assume that the configurations of $\theta$ are bounded by $\pm
\pi/2$.

The action $S_{\rm imp}(\theta)$ for the boundary degree of freedom is
\begin{equation}
S_{\rm imp}(\theta)=\int_{-\infty}^{+\infty} dx_2 \; \left[ {\frac{K}{2}}
 ({\frac{d \theta}{d x_2}})^2+ U(\theta(x_2))\right]
\label{eq:Simp}
\end{equation}
The full partition function which describes the quantum dynamics of both
fermions and the boundary degree of freedom is
\begin{equation}
{\cal Z}_{\rm full}=\int \; {\cal D}{\bar \psi} {\cal D}\psi {\cal D}\theta
\exp(-S_F({\bar \psi},\psi)-S_{\rm imp}(\theta))
\label{eq:PF}
\end{equation}
where $S_F({\bar \psi},\psi)$ is given in Eq.~(\ref{eq:fermiaction1}).
The coupling between the fermions (interacting or not) and the quantum
dot is implemented through the dynamical boundary condition at the edge $x_1=0$
\begin{equation}
R(0,x_0)=-e^{2i\theta(x_0)} L(0,x_0)
\label{eq:BC}
\end{equation}
suitably analytically continued to imaginary time as discussed in
Section ~\ref{sec:Fermions}. The partition function of
Eq.~(\ref{eq:PF}) has the general qualitative features of the systems 
discussed by Caldeira and Leggett \cite{caldeira} in their
work on Macroscopic Quantum Coherence.

\section{Charge Fluctuations and Boundary States}
\label{sec:response}

In this section we will use the model and the methods presented in the
past sections to describe the charge fluctuations near the edge of the
quantum wire. We will use the standard linear response approach and
compute the fermionic current as a response to an external field. 
We will begin by considering the effects of classical impurities and 
boundary degrees of freedom. Later in this section we will
also discuss the case of quantum impurites.

The expectation values of the fermionic current when
an external field ${\cal A}_\mu$ acts on the system is defined as
\begin{equation}
<J_{\mu}(x)>=-{1\over{{\cal Z}[{\cal A}]}}\frac{\delta {\cal Z}[{\cal
A}]}{\delta {\cal A}_\mu(x)}
\label{eq:currentdef}
\end{equation}
Since, as we have shown in section \ref{sec:boundary}, we have to
split the external
field ${\cal A}_{\mu}$ into two fields, $B_\mu$ and ${\bar b}_\mu$, we
define two currents, one in the bulk
\begin{equation}
<J_{\mu}(x)>=-{1\over{{\cal Z}[B_\mu,{\bar b}_\mu]}}\frac{\delta {\cal
Z}[B_\mu,{\bar b}_\mu]}{\delta
B_\mu(x)}
\label{eq:currentdefbulk}
\end{equation}
for $x_1>0$ and the other at the boundary
\begin{equation}
<J_{\mu}(0,x_2)>=-{1\over{{\cal Z}[B_\mu,{\bar b}_\mu]}}\frac{\delta {\cal
Z}[B_\mu,b_\mu]}{\delta {\bar b}_\mu(x_2)}
\label{eq:currentdefbundary}
\end{equation}
Using  Eq.~(\ref{eq:pfclasica}) we obtain
\begin{equation}
<J_{\mu}(x)>=\frac{1}{(\pi+ g^2)}\int_{\Omega} d^2y
{\Gamma}_{\mu \nu}(x,y) (B_\nu+{\bar A}^{\rm imp}_\nu+{\bar b}_\nu)(y)
\label{eq:jotamu}
\end{equation}
in the bulk, and
\begin{eqnarray}
<J_2(0,x_2)>&=&\frac{1}{(\pi+ g^2)}\int_{\Omega} d^2y {\Gamma}_{2\mu }(x_2,y)
B_\nu(y)\nonumber\\ &&+{1\over {\pi}} \int dy_2 K(x_2,y_2)
\left[\theta+\frac{\pi}{(\pi+ g^2)}({\bar A}^{\rm imp}_2+{\bar
b}_2)\right](y_2)
\label{eq:jota2}
\end{eqnarray}
at the boundary. The spatial component of the current at the boundary
vanishes because $\Gamma_{1\nu}(x,y) |_{x_1=0} = 0$.

It is interesting to note that in the expressions above $\theta$ and 
${\bar A}^{\rm imp}_2$ play different roles. While $\theta$ is
never multiplied by the coupling constant $g$, the factor in front of
${\bar A}^{\rm imp}_2$ is $\frac{\pi}{(\pi+ g^2)}$. That is, $\theta$
is not affected by the electron-electron interactions inside the wire.
We can understand this by recalling that $\theta$ represents the phase
shift due to the quantum dot. Physically, the electrons inside the
wire cannot screen the potential of the dot which lays outside the
system. In contrast, a local imperfection of the wire ,represented by
${\bar A}^{\rm imp}_2$ will see the effects of screening and hence its
effects will be affected by the electron-electron interactions. Recall
that in the calculation of the fermion determinant in reference 
\cite{paper1} the dynamical degree of freedom $\theta$ was placed
exactly at the origen, while ${\bar A}^{\rm imp}_2$ was placed at
$x_1=\epsilon$, that is ``in the bulk".  That means that $\theta$ is 
an ``external" boundary condition and ${\bar A}^{\rm imp}_2$ acts as 
an ``internal" perturbation to the system, affected by screening effects.
There are similar issues involving the current and charge distributions
as well. In fact, 
Maslov and Stone\cite{maslov} have reexamined recently the computation of
the conductance in a Luttinger liquid. They showed that the renormalization 
of the conductance\cite{kane1} $G={\frac{e^2}{h}}/(1+{\frac{g^2}{\pi}})$ 
(implied by the bulk bosonization equation~\ref{eq:bose-identity}) is
absent if the conductance is measured from outside the system ({\it i.~e\/},
in the external leads). 

It is instructive to calculate the induced total charge and to determine
the effects of the interactions on this quantity. As it turns out, the
induced charge depends on whether the boundary degree of freedom
$\theta$  is quantum mechanical or not. 

We first consider the classical case. We set $B_\mu={\bar b}_{\mu}=0$,
and calculate $Q_{\rm ind} (x_2)= \int_0^{\infty} dx_1 J_2(x_1,x_2)$.
Using Eqs.~(\ref{eq:jotamu}) and (\ref{eq:jota2}) the induced charge is
\begin{eqnarray}
Q_{\rm ind} (x_2)&=&{1\over {\pi}} \int _0^a dx_1 \int_{-\infty}^{+\infty}
 dy_2 K(x_2,y_2)
\left(\theta+\frac{\pi}{(\pi+ g^2)}{\bar A}^{\rm imp}_2 \right)(y_2)
\nonumber\\
&& + \frac{1}{(\pi+ g^2)} \int _a^{\infty} dx_1 \int_{\Omega} d^2y
{\Gamma}_{2 \nu}(x,y) {\bar A}^{\rm imp}_\nu(y)
\label{eq:curr1}
\end{eqnarray}
where the spatial integration is over the right half-plane.
Recalling that $A_{\mu}^{\rm imp}(x)= \delta(x_1) {\bar
A}_{\mu}(x_2)$, and that $\Gamma_{21}(x,y)|_{y_1=0}=0$ we obtain
\begin{eqnarray}
Q_{\rm ind}(x_2)&=& {a\over \pi} \int_{-\infty}^{+\infty}
 dy_2 K(x_2,y_2)\; \theta(y_2)
\nonumber\\
&+& {1\over {(\pi + g^2)}} \int _0^a dx_1  \int_{-\infty}^{+\infty}
 dy_2 K(x_2,y_2)
{\bar A}^{\rm imp}_2 (y_2) \nonumber\\
&& +\frac{1}{(\pi+ g^2)}  \int _a^{\infty} dx_1 \int_{-\infty}^{+\infty}
 dy_2
{\Gamma}_{22}(x,y_2) {\bar A}^{\rm imp}_2(y_2)
\label{eq:curr2}
\end{eqnarray}
Since the  kernel $K(x_2,y_2)$ can be written as $K(x_2,y_2)= \lim_{x_1
\rightarrow 0} \Gamma_{22}(x,y_2)$, it is easy to see that
the second and the third terms in Eq.~(\ref{eq:curr2}) cancel each
other, and the induced charge results
\begin{equation}
Q_{\rm ind}={a \over {\pi}}\int_{-\infty}^{+\infty}
 dy_2\; K(x_2-y_2) \; \theta (y_2)
\label{eq:101}
\end{equation}
For a static boundary degree of freedom, $\theta={\rm constant}$~,
the last equation gives $Q_{\rm ind}={\theta \over {\pi}}$ which 
is the standard formula for the charge fractionalization in the 
semiclassical theory of solitons \cite{yamagishi,wilczek,ruso}.
It is worth to emphasize that $Q_{\rm ind}$ is not due to a
non-conservation of charge in the wire. Instead, we will view the 
induced charge as resulting from the fermions being pushed in and out 
of the wire, back and forth from  the state that leaks to the left
of $x_1=0$, by the fluctuations of the quantum dot. The aparently
missing (or excess) charge resulting from this mechanism is
compensated  by fluctuations of opposite sign at infinity (the other 
end of the wire). In our construction of the quantum dot
coupled to the quantum wire of section \ref{sec:boundary} we found that,
in the abscence of tunneling between the dot and the wire, $\theta$ 
was restricted to the interval $(-{\frac{\pi}{2}},{\frac{\pi}{2}}]$. 
Hence, the total induced charge $Q_{\rm ind}$ takes values in the
range  $- {\frac{1}{2}}<Q_{\rm ind} \leq {\frac{1}{2}}$.
The ``extra state" implied by the leaking of the wave functions of the wire
into the regions where the electrons are strongly repelled by the dot is
mathematically analogous to the left ``half" of the midgap state of a 
soliton\cite{jackiw}. Notice that if $\theta$ is allowed to wander 
beyond its natural range shifted by  $N \pi$ (with $N$ an integer), 
the induced charge would instead be shifted by $N$. Such an effect can
only happen in the presence of real tunneling between the wire and the dot.

We consider now $\theta$ as a quantum degree of freedom with the dynamics
discussed in the previous section. From the expresion for 
$S_{\rm imp}(\theta)$ it is clear that the inertial term acts like a
high frequency cutoff.  Therefore, at low frequencies the induced term
in the action are dominant and we can neglect the
kinetic term. We should (and do) keep the potential term. We will only
discuss here the semiclassical approximation to this path-integral.
In this approximation the integration range can be extended to infinity if the
potential is present. In this limit, the potential can be replaced by
a linearized approximation of the form  $U( \theta)\approx  G
\theta^2$. Notice, however that as $G \to 0$ the issue of the
integration range becomes important once again.

In this limit, using Eq.~(\ref{eq:ifclass}), integrating $\theta$ we get
\begin{eqnarray}
{\cal Z}[B_\mu,b_\mu]={\cal K}\;\exp&&\left( {\frac{1}{2(g^2+\pi)}}
\int_\Omega d^2x \int_\Omega d^2y \;\; B_\mu (x){\Gamma}_{\mu \nu}(x,y)
B_\nu(y)\right.\nonumber\\
&&+\left.{\frac{1}{(g^2+\pi)}}
\int_\Omega d^2x\int_{-\infty}^{+\infty} dy_2 \;\;
B_\mu (x){\Gamma}_{\mu 2}(x,y_2)({\bar A}^{\rm
imp}_2+{\bar b}_2)(y_2)\right.\nonumber\\
&&-\left.\frac{g^2}{2\pi(\pi+ g^2)}\int_{-\infty}^{+\infty} \!dx_2\!\!
\int_{-\infty}^{+\infty} dy_2 ({\bar A}^{\rm
imp}_2+{\bar b}_2)(x_2)K(x_2,y_2)({\bar A}^{\rm imp}_2+{\bar b}_2)(y_2)
\right. \nonumber\\
&&+\left. {G \over 2} \int_{-\infty}^{+\infty} dx_2
({\bar A}^{\rm imp}_2+{\bar b}_2)^2(x_2)
\right. \nonumber\\
&& - \left. {G^2 \over 2} \int_{-\infty}^{+\infty} \!dx_2
\int_{-\infty}^{+\infty}  \! dy_2 ({\bar A}^{\rm imp}_2+{\bar b}_2)(x_2)
{\tilde K}(x_2 -y_2)  ({\bar A}^{\rm imp}_2+{\bar b}_2)(y_2)\right)
\end{eqnarray}
In this approximation, the kernel ${\tilde K}(x_2 -y_2)$ is
\begin{equation}
{\tilde K}(x_2 -y_2)=  e^{\pi G \mu} {\rm E}_1 (\pi G \mu) +
 e^{\pi G \mu ^*} {\rm E}_1 (\pi G \mu^*)$, and $\mu= a +i(x_2-y_2)
\label{eq:tilde}
\end{equation}
where ${\rm E}_1(x)$ is the exponential integral function.
In the bulk, the induced currents now take the form
\begin{equation}
<J_{\mu}(x)>=\frac{1}{(\pi+ g^2)}\int_{\Omega} d^2y
{\Gamma}_{\mu \nu}(x,y) (B_\nu(y) +{\bar A}^{\rm imp}_\nu+{\bar b}_{\nu})(y)
\end{equation}
At the boundary we find instead
\begin{eqnarray}
<J_2(x_2)>=&&\frac{1}{(\pi+ g^2)}\int_{\Omega} d^2y
{\Gamma}_{2 \nu}(x_2,y) B_\nu(y) - \frac{g^2}{\pi(\pi+ g^2)}
\int_{-\infty}^{+\infty} dy_2
K(x_2,y_2)({\bar A}^{\rm imp}_2+{\bar b}_2)(y_2) \nonumber\\
&&+G  ({\bar A}^{\rm imp}_2+{\bar b}_2)(x_2)
 -  G^2 \int_{-\infty}^{+\infty} dy_2 {\tilde K}(x_2 -y_2)
({\bar A}^{\rm imp}_2+{\bar b}_2)(y_2)
\end{eqnarray}
To compute the charge induced by the impurity we set $B_\mu={\bar
b}_2=0$. Therefore
\begin{eqnarray}
Q_{\rm ind}(x_2)&&\equiv \int_0^a dx_1<J_{2}(x_2)> + \int _a^\infty
dx_1<J_{2}(x_1,x_2)>\nonumber\\
&=& - \frac{g^2}{\pi(\pi+ g^2)} \int_0^a dx_1 \int_{-\infty}^{+\infty} dy_2
K(x_2,y_2){\bar A}^{\rm imp}_2(y_2) +  G \int_0^a  dx_1
({\bar A}^{\rm imp}_2+{\bar b}_2)(x_2) \nonumber\\
&& -  G^2 \int_0^a dx_1  \int_{-\infty}^{+\infty} dy_2 \; {\tilde K}(x_2 -y_2)
({\bar A}^{\rm imp}_2+{\bar b}_2)(y_2)  \nonumber\\
&+&\frac{1}{(\pi+ g^2)} \int_a^{\infty} dx_1 \int_{-\infty}^{+\infty} dy_2
{\Gamma}_{\mu 2}(x,y_2){\bar A}^{\rm imp}_2(y_2)
\end{eqnarray}
In this expression we see that
$\lim_{x_1\rightarrow 0}<J_{2}(x)>\neq <J_{\mu}(x_2)>$.
This discontinuity gives rise to the induced charge.  Replacing
$K(x_2,y_2)= \lim_{x_1
\rightarrow 0} \Gamma_{22}(x,y_2)$, and integrating over $x_1$ we
can write
\begin{eqnarray}
Q_{\rm ind}(x_2)=&& \lim_{a \rightarrow 0} \left( {1 \over \pi^2}
\int_{-\infty}^{+\infty} dy_2 {\bar A}^{\rm imp}_2(y_2)
{a \over { a^2 + (x_2- y_2)^2}}
\right. \nonumber\\
&&+ \left. a G  ({\bar A}^{\rm imp}_2+{\bar b}_2)(x_2)
 -  a G^2  \int_{-\infty}^{+\infty} dy_2 \; {\tilde K}(x_2 -y_2)
({\bar A}^{\rm imp}_2+{\bar b}_2)(y_2) \right)
\end{eqnarray}
According to the regularization procedure that has been used
throughout this work, $ \lim_{a \rightarrow 0} {a \over
{ a^2 + (x_2-y_2)^2}} = \pi \delta(x_2 - y_2)$. Therefore, after
taking the limit $a\rightarrow 0$ the induced charge results
\begin{equation}
Q_{\rm ind}(x_2)=\frac{1}{\pi}{\bar A}^{\rm imp}_2(x_2)
\end{equation}
It is clear that, once the boundary degree of freedom $\theta$ is
taken as a dynamical variable, the induced charge is independent of
the interaction. In other words,
there is no screening of the impurity due to the interactions.

\section{Bosonization of the fermi operators}
\label{sec:operators}

In this section we derive the one-particle Green's function from which
a set of bosonization rules for the fermion operators for an
interacting Fermi system with boundaries can be derived. For an
infinite system, these rules are the well known Mandelstam
formulas\cite{mandelstam}. In reference\cite{paper1} we derived a set of
rules for non-interacting fermions with boundaries. Here we will follow
the same methods and extract this information from the one-particle Green's
function, which is defined as
\begin{equation}
S^{\alpha\beta}(x,y)=\frac{\int {\cal D} {\bar \psi} {\cal D} \psi
\;{\cal D} a_\mu\;{\cal D}{\psi}_\alpha(x){\bar \psi}_\beta(y)\;
e^{-S_F(\bar \psi,\psi,a_\mu)}}{\int
{\cal D} {\bar \psi} {\cal D} \psi \;{\cal D} a_\mu\;e^{-S_F(\bar
\psi,\psi,a_\mu)}}
\label{teto}
\end{equation}
 with $\alpha,\beta=1,2$. $S_F$ is the action defined in
Eq.~(\ref{eq:fermiaction1}) with ${\cal A}_\mu$ set to zero ,{\it i.~e.\/} ,
with no external sources. That is, in Eq.~(\ref{teto}),
\begin{equation}
S_F ({\bar \psi},\psi,a_\mu)=\int_{\Omega} d^2x \;
{\bar \psi}( i \slp + \sla +\slimpA )\psi +\int_{\Omega} d^2x \;
{a_\mu^2 \over 2g^2}
\label{eq:fermiaction2}
 \end{equation}
We call ${\cal Z}^{\alpha\beta}_{\rm num}$ the numerator of
$S^{\alpha\beta}(x,y)$. That is
\begin{equation}
{\cal Z}^{\alpha\beta}_{\rm num}\equiv\int {\cal D} {\bar \psi} {\cal
D} \psi \;{\cal D} a_\mu\;{\psi}_\alpha(x){\bar
\psi}_\beta(y)\;e^{-S_F(\bar
\psi,\psi,a_\mu)}
\end{equation}
 In order to compute ${\cal Z}^{\alpha\beta}_{\rm num}$ we decouple
the fermion from the bulk H-S field, $a_\mu$, throught the gauge and
chiral transformations given by Eq.~(\ref{eq:chiral}). After the
transformation, the partition function becomes
\begin{eqnarray}
{\cal Z}^{\alpha\beta}_{\rm num}= {\cal K} &&\; {\rm Det} ( i \slp +
\slimpA)\;S^{\alpha\beta}_F(x,y|\theta,{\bar A}^{\rm imp})\nonumber\\
&&\int {\cal D}\eta\; \exp\left( -{1\over 2g^2}\int (\partial_\mu\eta)^2 +
\eta(x) -
\eta(y)\right)\nonumber\\
&&\int {\cal D}\phi\; \exp\left( -{1\over 2}({1\over \pi}+{1\over g^2})\int
(\partial_\mu\phi)^2
\right.\nonumber\\
&&\;\;\;\;\;\;\;\;\;\left.+{1\over \pi}\int \epsilon_{\mu\nu} {\bar A}^{\rm
imp}_\mu \partial_\nu\phi+
\sigma_\alpha \phi(x) + \sigma_\beta\phi(y)\right)
\label{potoden}
\end{eqnarray}
where $\sigma_1=1$ and $\sigma_2=-1$ and the indices ${\alpha ,\beta}$
are not contracted. The one-particle Green's function that appears in
Eq.~(\ref{potoden}) is, by definition
\begin{equation}
S^{\alpha\beta}_F(x,y|\theta,{\bar A}^{\rm imp})=\frac{\int {\cal D}
{\bar \chi} {\cal D}
\chi\; \chi_\alpha(x){\bar \chi}_\beta(y)\; \exp\{ \int{\bar \chi}( i
\slp +  \slimpA )\chi \}}
{\int {\cal D} {\bar \chi} {\cal D} \chi\; \exp\{ \int{\bar \chi}( i \slp
+ \slimpA )\chi \}}
\end{equation}
Note that
\begin{equation}
{\rm Det} ( i \slp +  \slimpA)=
{\int {\cal D} {\bar \chi} {\cal D} \chi\; \exp\{ \int{\bar \chi}( i \slp
+ \slimpA )\chi \}}
\end{equation}
The integations over $\eta$ and $\phi$ in Eq.~(\ref{potoden})are
straightfordward and give
\begin{equation}
I_\eta=\exp\left({g^2\over 2}\int_\Omega d^2z\int_\Omega d^2\;
wJ(z)G_N(z,w) J(w)\right)
\label{inteta}
\end{equation}
and
\begin{eqnarray}
I_\phi=\exp\left(-\frac{\pi}{2({\pi \over g^2} +1)}\int_\Omega d^2z\int_\Omega
d^2w\right.&&\left.[J_{\alpha\beta}(z)-
{1\over \pi}\epsilon_{\mu\nu}\partial_\nu( {\bar A}^{\rm imp}_\mu)(z)]
G_D(z,w)\right.\nonumber\\
&&\left.[J_{\alpha\beta}(w)-
{1\over \pi}\epsilon_{\mu\nu}\partial_\nu( {\bar A}^{\rm imp}_\mu)(w)]\right)
\label{intphi}
\end{eqnarray}
with $J(z)= \delta(z-x)-\delta(z-y)$ and
$J_{\alpha\beta}(z)=\sigma_\alpha \delta(z-
x)+\sigma_\beta\delta(z-y)$.
The function  $G_N(x,y)$ is the Von Newman Green's function in $\Omega$ which
satisfies
\begin{equation}
 \left\{
\begin{array}{ll}
\partial ^2_x G_N(x,y)= \delta(x-y)\\
\partial_{1}G_N(x_1,x_2;y)|_{x_1=0}=0
\end{array}
\right.
\label {eq:GN}
\end{equation}
The solution of Eq.~(\ref{eq:GN}) is
\begin{equation}
G_N(x,y) = {1\over {4\pi}} (\ln { {(x_1 - y_1)^2 + (x_2 - y_2)^2 +
a^2}\over a^2} + \ln {{(x_1 + y_1)^2 + (x_2 - y_2)^2 + a^2}\over a^2 })
\label{eq:GNE}
\end{equation}
where $a$ is the short distance cut-off.

Since the fermionic fields ${\bar \chi}$ and $\chi$ satisfy ${\bar
\chi}_R(x_2)=- e^{2\theta(x_2)}\chi_L(x_2)$, the ${\cal D}et ( i \slp
+\slimpA)$ turns out to be (see \cite{paper1})
\begin{equation}
{\rm Det} ( i \slp  + \slimpA)=\exp\left(\frac{1}{2\pi}
\int_{-\infty}^{+\infty}dz_2 \int_{-\infty}^{+\infty} dw_2
(\theta+ {\bar A}^{\rm imp}_2)(z_2)K(z_2,w_2)(\theta+ {\bar A}^{\rm
imp}_2)(w_2)\right)
\end{equation}
In Appendix \ref{sec:AA} we calculate the one particle Green's function
$S_F(x,y|\theta,{\bar A}^{\rm imp})$. There we find that it has the form
\begin{eqnarray}
S_F^{\alpha\beta}(x,y|\theta,{\bar A}^{\rm imp}_\mu)&=&
\exp\left({1\over{\pi}}\int_{-\infty}^{+\infty}\;dz_2\;
{\bar A}^{\rm imp}_2 (z_2) \frac
{\sigma_\alpha x_1 +i(x_2-z_2)}{x_1^2 + (x_2-z_2)^2}\right)\nonumber\\
S_F^{0,\alpha \beta}(x,y|\theta) &&\exp\left({1\over{\pi}}
\int_{-\infty}^{+\infty}\;dz_2\;
{\bar A}^{\rm imp}_2(z_2) \frac {\sigma_\beta y_1 -i(y_2-z_2)}{y_1^2 +
(y_2-z_2)^2}\right)
\end{eqnarray}
where $S_F^{0,\alpha \beta}(x,y|\theta)$ is the one particule Green's
function for free fermions in the half line with
$R(x_2)=-e^{2\theta(x_2)}L(x_2)$ which was calculated in
\cite{paper1}.

In order to simplify the notation, we define $R_{\alpha\beta}(w_2)$
and $R(w_2)$ as
\begin{equation}
R_{\alpha\beta}(w_2)=
{1\over{\pi}}\int_{-\infty}^{+\infty}\;dz\; \frac{z_1}{z_1^2 + (w_2-
z_2)^2}J_{\alpha\beta}(z)
\end{equation}
and
\begin{equation}
R(x_2)={1\over{\pi}}\int_{-\infty}^{+\infty}\;dz\; \frac{z_1}{z_1^2 +
(z_2-x_2)^2}J(z)
\end{equation}
Using these results, the one particle Green's function of Eq.~(\ref{teto})
can be shown to have the form
\begin{eqnarray}
S^{\alpha\beta}(x,y)&=&S_F^{0,\alpha \beta} (x,y|\theta)\;\;
\exp\left(\int_{z_2}\left[ R_{\alpha\beta}(z_2) \frac{\pi}{(\pi + g^2)}
{\bar A}^{\rm imp}_2(z_2)
+i R(z_2) {\bar A}^{\rm imp}_2(z_2) \right]\right)
\nonumber\\
&&
\exp\left(\int_{z,w}\left[ {g^2\over 2} J(z)G_N(z,w)
J(w)-\frac{\pi}{2({\pi \over g^2} +1)}
J_{\alpha\beta}(z)G_D(z,w)J_{\alpha\beta}(w)\right]\right)\nonumber\\
&&
\end{eqnarray}
We can rewrite this one particule Green's function in a more
familiar way as
\begin{equation}
S^{\alpha\beta}(x,y) \:
 ={\frac{1}{2\pi}}\left(\frac{4x_1^2+a^2}{a^2}\right)^{{\Delta_1 \over
4}}\left(\frac{4y_1^2+a^2}{a^2}\right)^{{\Delta_1 \over 4}}
 \left(
\begin{array}{cc}
\frac{H^{11}}{{\bar w}|w|^{\Delta_2}|z|^{\Delta_1}}
& \frac{H^{12}}{z |z|^{\Delta_2}|w|^{\Delta_1}}\\
-\frac{H^{21}}{{\bar z}|z|^{\Delta_2}|w|^{\Delta_1}}
& -\frac{H^{22}}{w |w|^{\Delta_2} |z|^{\Delta_1}}
\end{array}
\;\; \right)
\label{eq:potoint}
\end{equation}
where ${\Delta_1}=\frac{g^2(2\pi+g^2)}{2\pi(\pi+g^2)}$ ,
${\Delta_2}=\frac{g^4}{2\pi(\pi+g^2)}$,
$z=(x_1-y_1)+ i(x_2-y_2)$ ,$w=(x_1+y_1)+ i(x_2-y_2)$ and
$H^{\alpha\beta}(x,y)$ is
\begin{eqnarray}
H^{\alpha\beta}(x,y)&=&\exp\left({i\over {\pi}}
\int_{-\infty}^{+\infty} \;dz_2 \; (\theta + {\bar
A}^{\rm imp}_2)(z_2)
\left[\frac{x_2-z_2}{x_1^2+(x_2-z_2)^2}-\frac{y_2-z_2}{y_1^2+(y_2-
z_2)^2}\right]\right.\nonumber\\
+&&\left.{1\over {\pi}}\int_{-\infty}^{+\infty} \; dz_2 \;
(\theta +{\pi \over {\pi+g^2}} {\bar A}^{\rm imp}_2)(z_2)
\left[\sigma_\alpha\frac{x_1}{x_1^2+(x_2-z_2)^2}+
\sigma_\beta\frac{y_1}{y_1^2+(y_2-z_2)^2}\right]\right)
\end{eqnarray}
These results coincide with the one obtained by Fabrizio {\it et.~al.\/}
\cite{fabrizio}(in the limit of $L\rightarrow \infty$ and for spinless
fermions in that reference ) if we set $\theta = A^{\rm imp}_{\mu}=0$.

It is interesting to analize the behavior of the one particle
Green's function away and close to the boundary.  The condition on $z$
and $w$ to be far from the boundary is given by $|x_1-y_1| \ll \min
(x_1,y_1)$ and $(x_2-y_2) \ll \min (x_1,y_1)$ . This last condition
comes from the fact that if it does not hold, the excitations will
have time to reach the boundary and ``return" with information about
it to the points where the Green's function is beeing calculated.  In
this limit the diagonal terms of Eq~(\ref{eq:potoint}) vanish because
$\Delta_1-(\Delta_2 + 1) <0$ for any value of $g^2$. Then the one
particle Green's function becomes
\begin{equation}
S^{\alpha\beta}(x,y) \:  ={\frac{1}{2\pi}}
 \left(
\begin{array}{cc}
0&
\frac{H^{12}}{z |z|^{\Delta_2}}\\
-\frac{H^{21}}{{\bar z}|z|^{\Delta_2}}
& 0
\end{array}
\;\; \right)
\label{eq:potointlejos}
\end{equation}
which is the one particle Green's function for the whole line if we
take $\theta ={\bar A}_2=0$ in $H^{12}$ and $H^{21}$. Note also that
the diagonal terms vanish as $ x_1^{-\frac{\pi}{\pi
+g^2}}$. Therefore they go to zero slower than in the free
case. Physically this means that the presence of the boundary can be
felt further away when the electrons interact among themselves. That
is, the electron interaction ``propagate" the boundary into the bulk.
The condition for $z$ and $w$ to be close to the boundary is that
$x_1,y_1 \ll \min{x_2,y_2}$.
In this limit the components of $S^{\alpha\beta}(x,y)$ become
proportional to $|x_2-y_2|^{-({\frac{g^2}{\pi}}+1)}$.
Thus we see here the effect of the interactions is to change the
exponent of the power law at the boundary  relative to the free electron gas.
In particular, it goes to zero faster than in the free case.
We remark that the asymptotic behaviors for the one
particle Green's function, far away
and close to the boundary is in total agreement with results from
Conformal Field Theory\cite{Cardy}.

Up to this point we have considered the boundary condition $\theta$ as
a classical degree of freedom. To se the effect in the one particle
Green's function of a quantum degree of
freedom at the boundary we have to integrate over
$\theta$ in eq.~(\ref{eq:potoint}). We
follow the same steps described in reference \cite{paper1} to obtain
\begin{equation}
S^{\alpha\beta}(x,y) \:  ={\frac{1}{2\pi}}\left(\frac{(2x_1)^2+a^2}
{a^2}\right)^{{\Delta_1 \over
4}}\left(\frac{(2y_1)^2+a^2}{a^2}\right)^{{\Delta_1 \over 4}}
 \left(
\begin{array}{cc}
0
& -\frac{{\tilde H}^{12}}{z |z|^{\Delta_2}|w|^{\Delta_1}}\\
\frac{{\tilde H}^{21}}{{\bar z}|z|^{\Delta_2}|w|^{\Delta_1}}
&0
\end{array}
\;\; \right)
\label{eq:potointquant}
\end{equation}
where
\begin{eqnarray}
{\tilde H}^{\alpha\beta}(x,y)=&&-{\frac {g^2}{\pi(\pi+g^2)}}
\int_{-\infty}^{+\infty}\;dz_2
\left[\sigma_\alpha\frac{x_1}{x_1^2+(x_2-z_2)^2}+
\sigma_\beta\frac{y_1}{y_1^2+(y_2-z_2)^2}\right]
{\bar A}^{\rm imp}_2(z_2)\nonumber\\
&&
\end{eqnarray}
At this point, we make the analytic continuation ${\bar A}^{\rm imp}_2
\rightarrow i{\bar A}^{\rm imp}_0$ which set us back in Minkowski space
(real time). We
can see then that ${\tilde H}^{\alpha\beta}(x,y)$ becomes a pure
phase. Moreover, for a static impurity ,that is when ${\bar A}^{\rm
imp}_0$ is a constant, ${\tilde H}^{\alpha\beta}(x,y)=1$ for
$\alpha\neq \beta$. In other word, the phase shift for
$<\psi^\dagger_R\psi_R>$ and $<\psi^\dagger_L\psi_L>$ vanishes for
static impurities.  This result can be understood if we think in terms
of the analogous result in the whole line. It can be seen that for the
whole line and for a forward scattering static impurity ${\bar A}^{\rm
imp}_0$
\begin{equation}
<\psi^\dagger_R(x)\psi_R(y)>=
e^{-i\frac{{\bar A}^{\rm imp}_0}{1+ \frac{g^2}{\pi}}}\;\;
<\psi^\dagger_R(x)\psi_R(y)>|_{\rm Lutt}
, \;\;\;x_1>0,\;y_1<0
\end{equation}
 where $<\psi^\dagger_R(x)\psi_R(y)>|_{\rm Lutt}$ is the one particle
Green's function for the Luttinger liquid in absence of the
impurity. If both, $x_1$ and $y_1$ are either to the left or to the
right of the impurity, the phase factor vanishes. In the case we are
dealing with, $x_1$ and $y_1$ are always to the right of the impurity.
Hence, in the abscence of boundary backscattering terms,
the the only effect of the interactions is a
partial screening of the impurity represented by a factor of
$\frac{g^2}{\pi + g^2}$ in the phase shift.

\section{Conclusions}
\label{sec:conclusions}

In this paper we studied the Luttinger model in the half line
coupled to a quantum impurity at the origin. We generalized
the bosonization methods discussed in reference \cite{paper1} to
study the effects of interactions on fermi systems coupled to
impurities.  We gave special attention to the role of the fermion
boundary conditions on the bosonized effective theory. 
We showed that this model gives a good physical
picture of a quantum wire coupled with a quantum dot by a purely 
capacitive interaction. The fluctuations of the degree of freedom 
$\theta$ at the
boundary turn out to be  determined by the fluctuations of the charge 
on the quantum dot.
The screening effect at the boundary has been described.  We
calculated the induced
charge at the boundary in two cases, when the boundary condition is a
classical degree of freedom and when it is a quantum one. In the first
case, in accordance with semiclassical theories, we obtained charge
fractionalization. In the quantum case, the induced charge depends on
the impurity as expected.  
We also showed that the bosonized currents are modified by the
electron-electron interactions as well as by the presence of the
boundary degree of freedom, $\theta$. 
Finally we discussed the properties of the
one-particle green function (relevant for tunneling problems) far away
and close to the boundary.  

Our results in the case of open boundary
conditions, {\it i.e.}, for $\theta=0$,  coincide with the ones
obtained in the literature in that particular case (see for instance 
\cite{ejm}).
We stress here that with our approach it is possible
 to deal with a more general
case, since we can compute the
correlation functions in the case of arbitrary bounday conditions,
{\it i.~e.\/}, arbitrary $\theta$. In principle, by changing the dynamics
assigned to the boundary degree of freedom one should be able to study
different experimental realizations, like for instance a Luttinger
liquid coupled to a superconductor \cite{wfhs}. 
Moreover, this approach can be used to study other situations like
infinite systems coupled to impurities by using its representations as
a semi-infinite system coupled to an impurity at the origen with
properly defined boundary conditions \cite{thm}.

\section{Acknowledgements}

We are grateful to Matthew Fisher, Andreas Ludwig, Michael Stone and Amir
Yacoby for
many insightful comments. This work was supported in part
by the National Science Foundation through the grants NSF DMR94-24511 at
the University of Illinois at Urbana-Champaign,
NSF DMR-89-20538/24 at the Materials Research Laboratory of the University of
Illinois(MF,EF), by a Glasstone Research Fellowship in Sciences and a Wolfson
Junior
Research Fellowship (AL).

\newpage
\appendix
\section{Green's functions for Dirac Fermions coupled to singular gauge fields}
\label{sec:AA}

The aim of this section is to compute of the Green's function
\begin{equation}
S_F(x,y|\theta,{\bar A}^{\rm imp})=\frac{\int {\cal D} {\bar \chi} {\cal D}
\chi\;
\chi_\alpha(x){\bar \chi}_\beta(y)\; e^{ \int{\bar \chi}( i \slp  + \slimpA
)\chi }}
{\int {\cal D} {\bar \chi} {\cal D} \chi\; e^{ \int{\bar \chi}( i \slp
+ \slimpA )\chi}}
\label{verde1}
\end{equation}
By definition, this  Green's function satisfies the equation
\begin{equation}
(i\slp_x  + \slimpA)S_F(x,y|\theta,{\bar A}^{\rm imp})= \delta(x-y)
\end{equation}
What requieres a little more analysis is to determine the boundary condition
that this
Green's function satisfies at $x_1=0$.
At the begining, the functional space was the one whose functions at the
boundary
satisfy $R(x_2)=-e^{2\theta(x_2)}L(x_2)$. But as we showed in
reference\cite{paper1}, in
order to compute the determinant ${\cal D}et ( i \slp  + \slimpA)$ in this
space, it
is equivalent to work in the functional space whose functions satisfy, at the
boundary,
$R(x_2)=-e^{2(\theta+ {\bar A}^{\rm imp}_2)(x_2)}L(x_2)$ and compute on
such space, the determinant ${\cal D}et ( i \slp)$. Since  ${\cal D}et ( i \slp
+ \slimpA)$
is the denominator of Eq.~(\ref{verde1}), and since, obviously, the whole left
hand side of
Eq.~(\ref{verde1}) has to be computed consistently over a single functional
space,
the resulting Green's function is the Green's function over such functional
space.
Hence, the Green's function in Eq.~(\ref{verde1}) is the one over the
functional space
whose functions satisfy, at the boundary,$R(x_2)=-e^{2(\theta+{\bar A}^{\rm
imp}_2)(x_2)}L(x_2)$. Another way of understanding this point is recalling
that, in a
second quantification language, the denominator of Eq.~(\ref{verde1}) beeing a
normalization for the Green's function, defines the vacuum states. That is the
creation
and anihilation operators which expand the physical phase space. Hence if we
choose
such space as the one whose operators satisfy $R(x_2)=-e^{2(\theta + {\bar
A}^{\rm imp}_2)(x_2)}L(x_2)$, the boundary conditions for the  Green's
functions we
compute have the same boundary conditions.

Therefore, the aim now is to compute
\begin{equation}
 \left\{
\begin{array}{ll}
(i\slp_x + \slimpA)S_F(x,y|\theta,{\bar A}^{\rm imp})= \delta(x-y)\\
S^{11}_F(0,x_2;y|\theta,{\bar A}^{\rm imp})=-e^{2(\theta + {\bar A}^{\rm
imp}_2)(x_2)}S^{21}_F(0,x_2;y|\theta,{\bar A}^{\rm imp})
\end{array}
\right.
\label{eq:S}
\end{equation}
It is easy to see that if we choose $U$ and $V$ such that  ${\bar A}^{\rm
imp}_\mu(x_2)\delta(x_1)=\partial_\mu V-\epsilon_{\mu\nu}\partial_\nu U$ then
\begin{equation}
S_F(x,y|\theta,{\bar A}^{\rm
imp})=e^{iV(x)+\gamma_5U(x)}{S}^0_F(x,y|\theta,{\bar
A}^{\rm imp},U)e^{-iV(x)+\gamma_5U(x)}
\end{equation}
where $S^0_F(x,y|\theta,{\bar A}^{\rm imp},U)$ satisfies
\begin{equation}
 \left\{
\begin{array}{ll}
(i\slp_x ){S}^0_F(x,y|\theta,{\bar A}^{\rm imp},U)= \delta(x-y)\\
S_F^{0,11}(0,x_2;y|\theta,{\bar A}^{\rm imp},U)=-e^{2[(\theta + {\bar
A}^{\rm imp}_2)(x_2)- U(0,x_2)]}S^{0,21}_F(0,x_2;y|\theta,{\bar A}^{\rm imp},U)
\end{array}
\right.
\end{equation}
As has been shown in \cite{paper1},
\begin{eqnarray}
S^{0,\alpha\beta}_F(x,y|\theta,{\bar A}^{\rm imp},U)&=&
\exp\left({1\over{\pi}}\int
dz_2( {\bar A}^{\rm imp}_2 + \theta-U)(z_2) \frac {\sigma_\alpha x_1 +i(x_2-
z_2)}{x_1^2 + (x_2-z_2)^2}\right)\nonumber\\
S_F^{0,\alpha \beta} (x,y)&&\exp\left({1\over{\pi}}\int\;dz_2\; ({\bar A}^{\rm
imp}_2+\theta -U)(z_2) \frac {\sigma_\beta y_1 +i(y_2-z_2)}{y_1^2 +
(y_2-z_2)^2}\right)
\end{eqnarray}
 with $S_F^0(x,y)$ the Green's function
\begin{equation}
S_F^0(x,y) \:  ={\frac{1}{2\pi}}
 \left(
\begin{array}{cc}
-\frac{1}{(-x_1-y_1) + i(x_2-y_2)}
& -\frac{1}{(x_1-y_1)-i(x_2-y_2)}\\
\frac{1}{(x_1-y_1)+i(x_2-y_2)}
& \frac{1}{(-x_1-y_1) - i(x_2-y_2)}
\end{array}
\;\; \right)
\label{eq:poto}
\end{equation}
We have found that $U$ and $V$ above are
\begin{equation}
U(x)= {-1\over{\pi}}\int\;dz_2\; {\bar A}^{\rm imp}_1(z_2) \frac {x_2-
z_2}{x_1^2 + (x_2-z_2)^2} +{1\over{\pi}}\int\;dz_2\; {\bar A}^{\rm imp}_2(z_2)
\frac {x_1}{x_1^2 + (x_2-z_2)^2}
\end{equation}
and
\begin{equation}
V(x)= {1\over{\pi}}\int\;dz_2\; {\bar A}^{\rm imp}_1(z_2) \frac {x_1}{x_1^2 +
(x_2-z_2)^2} +
{1\over{\pi}}\int\;dz_2\; {\bar A}^{\rm imp}_2(z_2) \frac {x_2-z_2}{x_1^2 +
(x_2-z_2)^2}
\end{equation}
Using the results above we obtain
\begin{eqnarray}
S_F^{\alpha \beta}(x,y|\theta,{\bar A}^{\rm imp})&=&
\exp\left({1\over{\pi}}\int\;dz_2\; ( 2{\bar A}^{\rm imp}_2+ \theta -U)(z_2)
\frac
{\sigma_\alpha x_1 +i(x_2-z_2)}{x_1^2 + (x_2-z_2)^2}\right)\nonumber\\
&& \exp\left({1\over{\pi}}\int\;dz_2\; ( {\bar A}^{\rm imp}_1)(z_2) \frac {i
x_1 -
\sigma_\alpha(x_2-z_2)}{x_1^2 + (x_2-z_2)^2}\right)
S_F^{0,\alpha \beta}(x,y)\nonumber\\
&&\exp\left({1\over{\pi}}\int\;dz_2\; ( 2{\bar A}^{\rm imp}_2+\theta -U)(z_2)
\frac {\sigma_\beta y_1 -i(y_2-z_2)}{y_1^2 + (y_2-z_2)^2}\right)\nonumber\\
&&\exp\left({1\over{\pi}}\int\;dz_2\; ( {\bar A}^{\rm imp}_1)(z_2) \frac {-iy_1
-
\sigma_\beta (y_2-z_2)}{y_1^2 + (y_2-z_2)^2}\right)
\end{eqnarray}
It can be shown that
\begin{eqnarray}
{1\over{\pi}}\int\;dz_2\; U(z_2) \frac {\sigma_\alpha x_1 +i(x_2-z_2)}{x_1^2 +
(x_2-
z_2)^2}&=&{1\over{\pi}}\int\;dz_2\; {\bar A}^{\rm imp}_2(z_2) \frac
{\sigma_\alpha x_1 +i(x_2-z_2)}{x_1^2 + (x_2-z_2)^2} \nonumber\\
&&+{1\over{\pi}}\int\;dz_2\; {\bar A}^{\rm imp}_1(z_2) \frac {ix_1 -
\sigma_\alpha (x_2-z_2)}{x_1^2 + (x_2-z_2)^2}
\end{eqnarray}
Hence the Green's function $S_F(x,y|\theta,{\bar A}^{\rm imp},{\bar a})$
becomes
\begin{eqnarray}
S_F^{\alpha\beta}(x,y|\theta,{\bar A}^{\rm imp})&=&
\exp\left({1\over{\pi}}\int\;dz_2\; ( {\bar A}^{\rm imp}_2 +\theta)(z_2) \frac
{\sigma_\alpha x_1 +i(x_2-z_2)}{x_1^2 + (x_2-z_2)^2}\right)S_F^{0,\alpha \beta}
(x,y)\nonumber\\
&&\exp\left({1\over{\pi}}\int\;dz_2\; ({\bar A}^{\rm imp}_2+\theta)(z_2) \frac
{\sigma_\beta y_1 -i(y_2-z_2)}{y_1^2 + (y_2-z_2)^2}\right)
\end{eqnarray}
Note that this expression can be factorized as
\begin{eqnarray}
S_F^{\alpha\beta}(x,y|\theta,{\bar A}^{\rm imp})&=&
\exp\left({1\over{\pi}}\int\;dz_2\;  {\bar A}^{\rm imp}_2 (z_2) \frac
{\sigma_\alpha x_1 +i(x_2-z_2)}{x_1^2 + (x_2-z_2)^2}\right)S_F^{0,\alpha
\beta}(x,y|\theta)
\nonumber\\
&&\exp\left({1\over{\pi}}\int\;dz_2\; {\bar A}^{\rm imp}_2(z_2) \frac
{\sigma_\beta y_1 -i(y_2-z_2)}{y_1^2 + (y_2-z_2)^2}\right)
\end{eqnarray}
 where $S_F^0(x,y|\theta)$  satisfies Eq.~(\ref{eq:S}) with $\theta$ in the
boundary
condition.

\end{document}